\documentclass[10pt]{article}

\usepackage{amsmath}
\usepackage{amssymb}

\usepackage{graphicx}

\usepackage{cite}
\usepackage{color} 
\usepackage{url}  
\usepackage{ifthen}  

\usepackage[labelfont=bf,labelsep=period,justification=raggedright]{caption}

\makeatletter
\renewcommand{\@biblabel}[1]{\quad#1.}
\makeatother

\date{}

\begin{document}

\begin{flushleft}
{\Large
\textbf{Resampling effects on significance analysis of network clustering and ranking}
}
\\
Atieh Mirshahvalad$^{1,\ast}$, 
Olivier H. Beauchesne$^{2}$, 
\'{E}ric Archambault$^{3}$,
Martin Rosvall$^{4}$
\\
\bf{1} Integrated Science Lab, Department of Physics, Ume{\aa} University, Ume{\aa}, Sweden
\\
\bf{2} Science-Metrix, Montr\'{e}al, Canada
\\
\bf{3} Science-Metrix, Montr\'{e}al, Canada
\\
\bf{4} Integrated Science Lab, Department of Physics, Ume{\aa} University, Ume{\aa}, Sweden
\\
$\ast$ E-mail: atieh.mirshahvalad@physics.umu.se
\end{flushleft}

\section*{Abstract}
Community detection helps us simplify the complex configuration of networks, but communities are reliable only if they are statistically significant.
To detect statistically significant communities, a common approach is to resample the original network and analyze the communities.
But resampling assumes independence between samples, while the components of a network are inherently dependent. 
Therefore, we must understand how breaking dependencies between resampled components affects the results of the significance analysis.
Here we use scientific communication as a model system to analyze this effect.
Our dataset includes citations among articles published in journals in the years 1984-2010.
We compare parametric resampling of citations with non-parametric article resampling.
While citation resampling breaks link dependencies, article resampling maintains such dependencies.
We find that citation resampling underestimates the variance of link weights. Moreover, this underestimation explains most of the differences in the significance analysis of ranking and clustering.
Therefore, when only link weights are available and article resampling is not an option, we suggest a simple parametric resampling scheme that generates link-weight variances close to the link-weight variances of article resampling.
Nevertheless, when we highlight and summarize important structural changes in science, the more dependencies we can maintain in the resampling scheme, the earlier we can predict structural change.

\section*{Introduction}

Researchers use network theory\cite{newman2010Intro} to better understand complex systems\cite{VespignaniNPhys2011,Jeong2000,Kleinberg:2000p5066,Milo2002} with many interacting components\cite{RevModPhys.74.47,newman2003structure,boccaletti06,Sales-Pardo25092007,ClausetEtAl2008a}. In network theory, there is great interest in detecting the tightly interconnected structural patterns of the network, so-called communities\cite{girvan2002community,radicchi2004defining,Newman,danon2006effect,blondel2008fast,hastings2006community,rosvall2007information,palla2005uncovering,YY_LC_nature2010,ISI:000238278400002,Fortunato201075}. 
Community detection helps us simplify the structure of the network because the communities often correspond to functional units of the system. 
However, communities are reliable only if they are statistically significant \cite{Spirin2003,PhysRevE.82.066106,PhysRevE.81.046110,10.1371journal.pone.0018961}.
Detecting statistically significant communities is possible when we have many instances of the network, because
we can first identify communities in each of the instances and then assess the significance of each community. 
But most often, we only have a single observation of the real network. To overcome this challenge and detect significant communities of real networks, we need a statistically sound procedure that generates instances of the single raw network.

A common approach to generating instances of the raw network is to use resampling techniques \cite{gfeller2005finding,karrer2008robustness,rosvall2010mapping,SCSN}.
The idea behind the resampling approach is fairly simple, since we can view a network as the aggregation of many natural events.
When resampling, we simply imitate the process of the network formation and generate various realizations of the raw network.
With numerous resampled networks, we can aggregate the community information and determine which communities of the raw network are significant and to what degree.
The catch, however, is that we must assume that the events that generate the observed network are independent. 
Therefore, it is important to raise the question: How much do the results of the significance analysis depend on the different assumptions about independent events? 
Specifically, how important are the link correlations in the resampling scheme?

When resampling weighted networks, the significance of communities depends not only on the weights of the links but also on their individual link-weight variances and their neighbor link-weight correlations across the resamples (two links are neighbors if they share a common node).
Here we aim to explore how much the link-weight variances and correlations in different resampling schemes affect the results of significance analysis for weighted, directed citation networks aggregated at the journal level.
In previous work, and with data limited to citation counts between journals, we used Poisson resampling without link-weight correlations to generate bootstrap networks\cite{rosvall2010mapping}. That is, independently from other links, we resampled the weight of each weighted directed link from a Poisson distribution with mean equal to the original link weight. 
This independent citation resampling is an oversimplification. Citations in the same article depend on each other and introduce correlations: Citations to articles published in the same journal introduce \emph{within-link correlations} that affect the link-weight variance of individual links. Citations to articles published in different journals introduce \emph{between-link correlations} that affect the interdependence of the weights of neighbor links.
With access to article-level data, we now can resample articles and maintain link correlations to better assess the significance of communities as well as journal rankings. At the same time, we can better understand the effects of eliminated link correlations in Poisson resampling.

Our dataset includes citations between scientific articles published in journals in all areas of science in the years 1984-2010. For a specific year, we can build a weighted, directed network of scientific journals in which the weight of each link between two journals \emph{A} and \emph{B} represents the number of times that articles published in journal \emph{A} cite articles published in journal \emph{B}. Because we are interested in the frontier of science, we only include citations to articles published no more than three years back in time. For example, in the 2009 data set, there are 961,542 scientific articles and 11,373 journals. This gives a citation network of journals with 11,373 nodes and 1,195,928 weighted, directed links.
As in many other citation networks, the degree distribution is skewed with a power-law exponent just below two.
We use the network from 2009 in most of our analysis, except when analyzing change over time.
Since science is continuously growing, the network from 2009 is the largest in our data set.

\section* {Materials and Methods}
\begin{figure}[!ht]
\begin{center}
\includegraphics[width=0.5 \columnwidth]{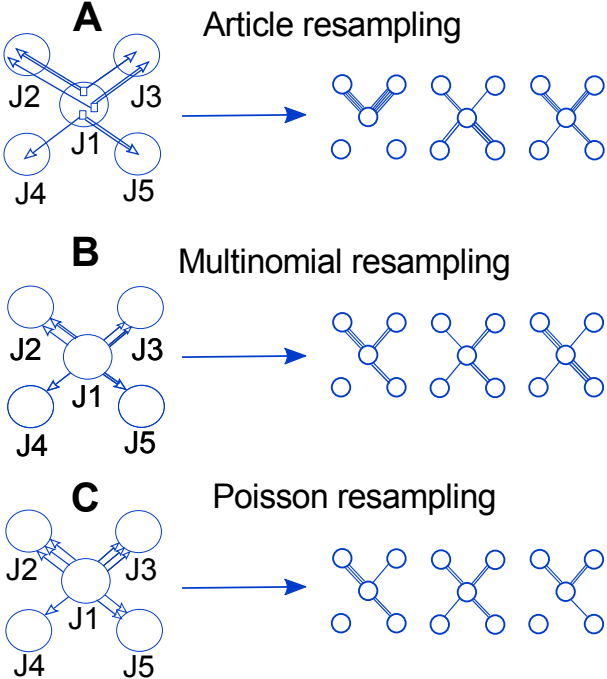}
\end{center}
\caption{
{\bf Link correlation preservation in different resampling schemes.} \textbf{A} Article resampling maintains correlations between links and also correlations within links. For example, an article in journal J1 might cite articles from journal J2 together with articles in journal J3 (correlations between links). An article in journal J1 might also cite another journal J2 more than once (correlations within links). The right-hand side shows some examples of possible resampled networks that necessarily keep correlation between and within links. \textbf{B} Multinomial resampling only maintains the correlations within links. The examples of resampled networks on the right-hand side show that they could be generated without keeping between-links correlations. \textbf{C} Poisson resampling does not maintain any link correlation. Every link of a resampled networks is generated independently of others.
}
\label{Schem}
\end{figure}

To understand what effect the link correlations of the resampling scheme have on assessing significant communities, we compared a resampling scheme that maintains between-link and within-link correlations (article resampling), a resampling scheme that only maintains within-link correlations (multinomial resampling), and a resampling scheme that maintains no link correlations (Poisson resampling), as shown in Fig.~\ref{Schem}.
In between-link correlations, the link weights of neighbor links are correlated. That is, in a resampled network, the weight of a link is not independent of the weight of a neighbor link. In within-link correlations, each link weight in a resampled network is the outcome of dependent events. Below 
we explain the three resampling methods: article resampling, multinomial resampling, and Poisson resampling. We also clarify the role of link-weight correlations in each method.
Article resampling is based on the assumption that articles can be treated independently of each other. That is, whether an article is published does not depend on whether other articles are published.
Assuming that we have a pool of all the articles that participate in our citation network, the process of article resampling to create bootstrap networks is simple.  
We randomly pick an article from the pool and add its citations from the journal in which the article was published to the cited journals.
Then we put this article back in the pool. We continue this process as many times as the number of articles in the original network. 
Since one article might cite articles in different journals,
article resampling automatically introduces correlations between the link-weights of the bootstrap networks. 
As Fig.~\ref{Schem}A shows, the links \emph{J1-J2} and \emph{J1-J3} are correlated because, for example, it is not possible to have a link \emph{J1-J2} and not a link \emph{J1-J3}. 

Article resampling also introduces within-link correlations, because an article might cite articles of a specific journal more than once.
In Fig.~\ref{Schem}A, for example, the link weight between \emph{J1} and \emph{J2} is three. This weight is not the outcome of three independent single citations, but rather is generated from one double and one single citation. Because the two citations in the double citation are dependent, and two, not three, events generated the link weight, the link variance will be higher over resampled networks than if the citations were sampled independently.
To investigate how these correlations affect the significance analysis, we compare article resampling with multinomial resampling, which keeps the correlations within link weights but destroys the correlations between link weights.

Multinomial resampling assumes that information about multiple citations from single articles to journals is known and can be treated independently.
To generate the bootstrap networks, we maintain the topology of the raw network and, independently for each link, resample its weight from a multinomial distribution with the set of multiple citations given by the article-level data. 
We emphasize that multinomial resampling does not maintain correlations between link weights, but it does maintain the correlations within link weights (Fig.~\ref{Schem}B). 
As a result, multinomial resampling creates an intermediate stage between a completely destroyed link correlation (Poisson resampling) and a fully maintained link correlation (article resampling). 
For example, in generating each link weight, multinomial resampling only includes the articles that contribute to that link weight and disregards other links that those articles might contain.
The question is: how much do the destroyed between-link correlations of multinomial resampling affect the significance analysis?
In section \emph{Results and Discussion}, we show that significant clusters generated with multinomial resampling are close to the significant clusters of article resampling. This result demonstrates that the role of between-link dependency on significance analysis of clusters is relatively small.

Poisson resampling assumes that citations can be treated independently of each other. The process of Poisson resampling for generating bootstrap networks is as follows:
we maintain the topology of the raw network and, independently for each link, resample its weight from a Poisson distribution with mean equal to the original link weight.
Poisson resampling not only automatically ignores the correlation between link weights, but also ignores the correlations within link weights (Fig.~\ref{Schem}C).
The question is: how much does the assumption about fully independent link weights affect the results of the significance analysis?
In section  \emph{Results and Discussion}, we show that Poisson resampling underestimates the variance of link weights compared to article resampling, and that within-link dependency does matter for the significance analysis of clustering and ranking. 

\section* {Results and Discussion}\label{ResDis}
In order to investigate the effect of link correlations on the significance analysis of clusters,
we create 1000 bootstrap networks based on a resampling scheme. 
Then we search for significant clusters, or cluster cores, which we define as the biggest subset of nodes in each cluster that gathered together in more than 90\% of the bootstrap networks. 
Correspondingly, a non-significant part of a cluster would be the subset of nodes in the cluster that is separated from the core in more than 10\% of bootstrap networks.
For clustering, we use \emph{infomap}, an information-theoretic algorithm that reveals regularities in a given network based on how information flows on that network \cite{Rosvall29012008}.
\begin{figure}[!ht]
\begin{center}
\includegraphics[width=0.5 \columnwidth]{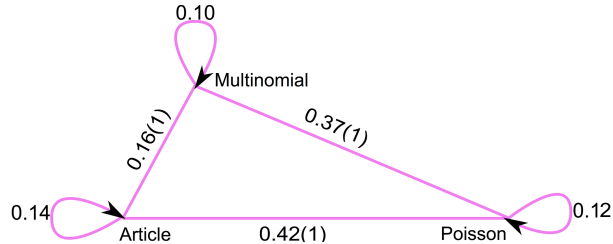}
\end{center}
\caption{
{\bf The differences between significant clusters' cores in different resampling schemes.}  We calculate normalized information distance ($d_{max}$) between the significant cores of the two corresponding methods with respect to the PageRank. All values correspond to an average over at least 2000 runs.
}
\label{CsigCFig}
\end{figure}

Figure \ref{CsigCFig} shows the difference between significant cluster cores of article, multinomial, and Poisson resampling in terms of \emph{normalized information distance}. The normalized information distance is defined as one minus the \emph{normalized mutual information}:
 \begin{equation}\label{Eqdmax}
 d_{max}=1-\frac{I(C,C')}{max(H(C),H(C'))},
 \end{equation} 
where $H(...)$ refers to Shannon entropy and $I(C,C')$ is the mutual information between the significant cores of the two resampling schemes that tells us how similar they are. Mutual information between two clusters $C$ and $C'$ is described as:  
\begin{equation} I(C;C') = \sum_{c,c'} P(c,c') \log{P(c,c') \over P(c) P(c')}
 \end{equation}
where $P(c,c') $ is the joint probability distribution between two clusterings $c$ and $c'$. $P(c)$ and $P(c')$ refer to the marginal probability distributions.

If $C$ and $C'$ are identical, then the normalized mutual information is equal to 1, which means that, by knowing one cluster structure, we know the other one. Conversely, if $C$ and $C'$ are completely independent, by knowing one, we learn nothing about the other one and the normalized mutual information between them would be 0.
We use normalized information distance for comparing clusterings because it is a sound metric\cite{Vinh2010ITM}.
Figure \ref{CsigCFig} shows that the difference between significant cores of article and multinomial resampling is of the same order as the difference between two iterations of each of these schemes, and both of them are considerably different from Poisson resampling. Although multinomial resampling does not hold the correlation between citations and article resampling does, our results show that between-link dependency does not have a great impact on the significance analysis of clusters.
\begin{figure}[!ht]
\begin{center}
\includegraphics[width=0.7\columnwidth]{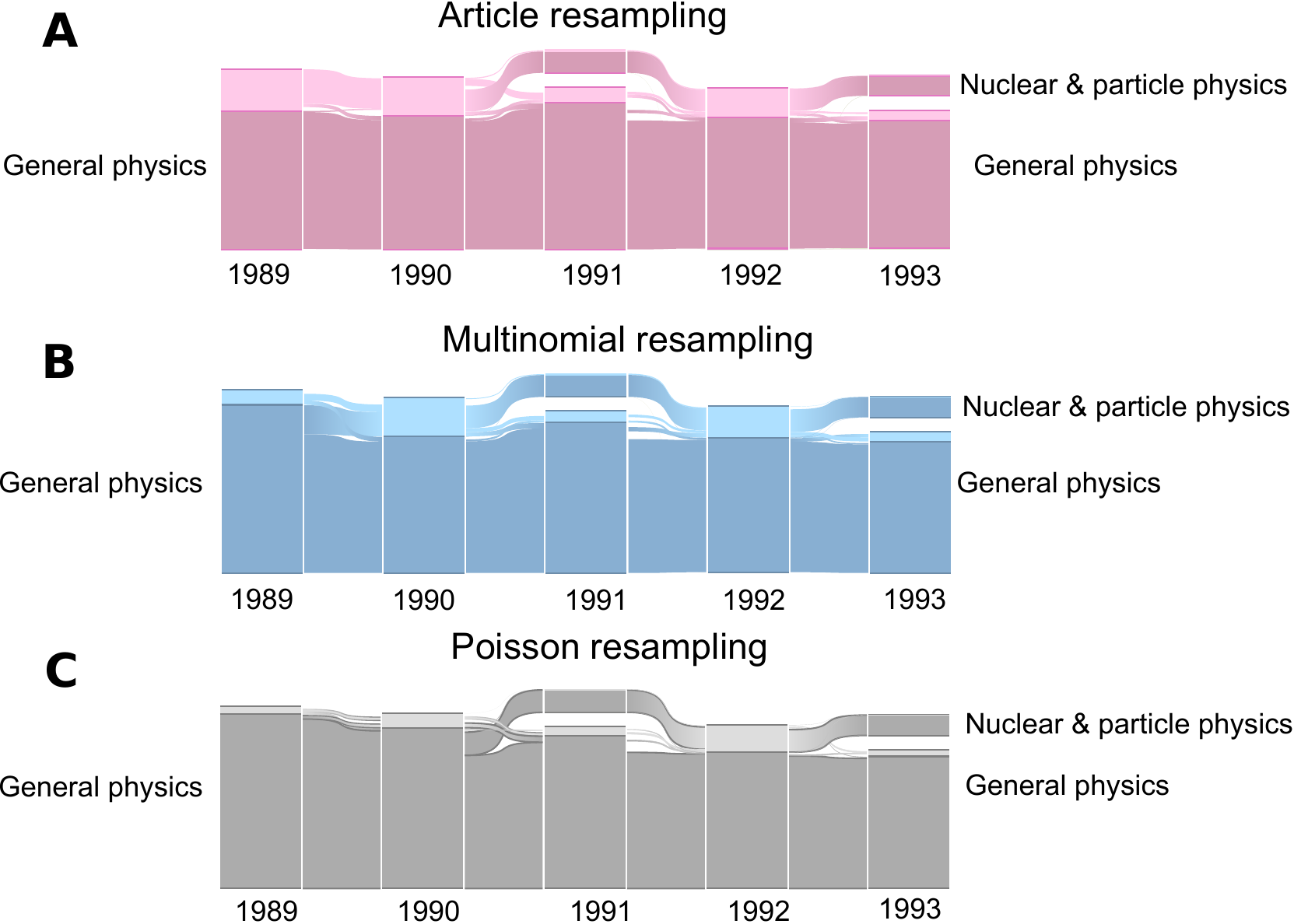}
\end{center}
\caption{
{\bf The separation of Nuclear \& particle physics from the Physics module. } 
In this diagram, each block in a given year corresponds to a specific module. 
In a block, the lighter colors represent the non-significant part of the module and the white vertical gap between blocks separates modules.
The stream field between two blocks in consecutive years shows changes that happen to a block.
While all three resampling schemes agree on the separation of \emph{Nuclear} \& \emph{particle physics} from \emph{General physics} into an independent stand-alone module by 1993,
article resampling emits a signal about this change sooner than multinomial or Poisson resampling.
}
\label{All1}
\end{figure}

We illustrate the effects of link-weight variance on clustering in a concrete example. Figure \ref{All1} shows the alluvial diagram of the three resampling schemes over the years 1989-1993.
Each block represents a specific module in a given year, and the height of a block represents its importance in terms of PageRank \cite{Brin1998107}. Based on the areas of specialization of the journals clustered together in each module, we manually label the modules.
In a block, the lighter colors correspond to the non-significant part of the module; the bigger this area is, the more non-significant nodes that module has. The white vertical gap between blocks separates the modules, and the numbers under each block correspond to the year. 
Blocks in a given year might merge as a single block in the next year, or a subset of a block might diverge from it in the next year. The changes that happen to a block from one year to the next are shown by the stream field between the two blocks.

As shown in the figure, all three resampling schemes agree on the separation of \emph{Nuclear} \& \emph{particle physics} from \emph{General physics} as an independent stand-alone module in 1993.
The exact year will depend on the citation window and data at hand, and by no means do we conclude that we see the emergence of a new field in 1993. While Nuclear \& particle physics was considered a research area long before 1993, it takes time before it shows up in the structure of the journal citation network.
Instead of singling out a particular year for the emergence of a scientific field, our main focus here is instead to show that different resampling schemes identify fields at different times (Fig.~\ref{All1}).
For example, in article resampling, the Nuclear \& particle physics module is highlighted as a non-significant part of General physics in 1989, while in Poisson and multinomial resampling, this happens later. 
In this way, the process of becoming non-significant could provide us a signal about important changes that might happen in the future; apparently, article resampling can give this signal sooner than multinomial resampling, and multinomial resampling can give it sooner than Poisson resampling. 
We conclude that for significance analysis of communities, within-link correlations play a more important role than between-link correlations.  
Moreover, maintaining link correlations in a resampling scheme can help us to identify the changes in a network earlier.

 \begin{figure}[!ht]
\begin{center}
\includegraphics[width=0.9 \columnwidth]{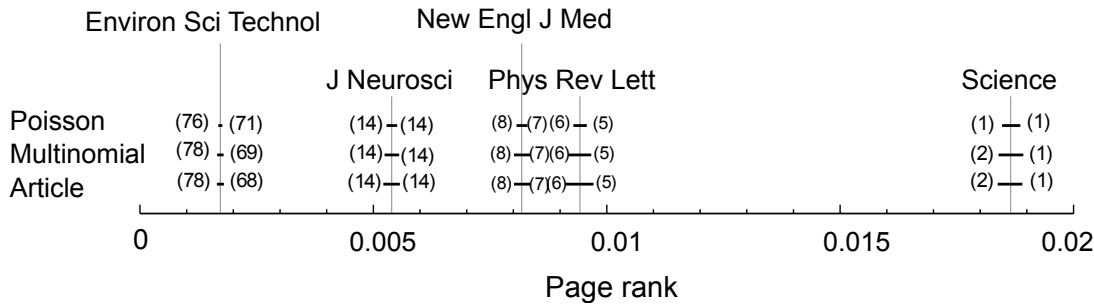}
\end{center}
\caption{
{\bf The variation of the PageRank for top-rank journals based on different resampling schemes.} In agreement with the result of single link-weight variance analysis, 
our analysis shows that core structures in article and multinomial resampling are much more similar to each other than in the Poisson resampling.
The article resampling is the biggest perturbation, in which the 95\% confidence interval for the PageRank is broader than in multinomial or article resampling. Multinomial and article resampling were second and third, respectively.}
\label{SigVarTop}
\end{figure}

As another example of significance analysis of an aggregated network measure, we analyze the effects of the different resampling schemes on PageRank.
In calculating PageRank, the importance of a node (a journal in our citation network) corresponds to the importance of nodes that cite this node, so the full network indirectly participates in calculating the PageRank of a node.
Figure \ref{SigVarTop} shows how much the PageRank of some top journals would vary based on the resampling scheme.
The length of each line corresponds to an interval that covers the variation of PageRank for a given journal in a given resampling scheme.
The numbers on the left/right hand side of each line correspond to the minimum/maximum rank order of each journal for a resampling scheme.
\emph{Science} has the largest PageRank value in the raw network, and so it is the first journal in the rank order. In Poisson resampling, \emph{Science} always maintains its first position in the ranking list.
But in multinomial and article resampling, \emph{Science} sometimes drops to the second position.
In a similar fashion, the rank order of PRL (\emph{Physical Review Letters}), NEJM (\emph{New England Journal of Medicine}), and J Neurosci (\emph{Journal of Neuroscience}) changes based on the resampling scheme that is used. In general, the PageRank of a node varies more in article resampling than in Poisson or multinomial resampling.
In this respect, we study the effect of resampling schemes on the rank order of all nodes in the network. 
We sample pairs of nodes $(i,j)$ from the rank order that we obtain from a resampling scheme and compare them with the rank order that we obtain from another resampling scheme.
We sample pairs of nodes proportional to their PageRank 
and measure the similarity between the two rank orders in terms of normalized mutual information. If, for all possible pairs in the two-rank order, the node with the highest rank in one order is the same in the other order, the mutual information between the two rank orders would be one.
The more different the two rank orders are, the smaller the mutual information between them would be. If the two rank orders do not have any common pair orders, the mutual information between them would be zero.
In a quantitative analysis of the rank order for the different resampling schemes, we find that the normalized information distance (Eq.~\ref{Eqdmax}) between two different rankings generated with the same resampling scheme is,
on average, about 26 percent larger for article resampling than for Poisson resampling and 23 percent larger for multinomial resampling than for Poisson resampling. For ranking, article resampling has the biggest variation, but multinomial resampling without correlations between links varies almost as much as article resampling. Multinomial resampling can explain almost all ranking variances of article resampling with correlations between links.

The between-link correlations of article resampling seem to play a minor role on significance analysis on ranking (Fig.~\ref{SigVarTop}) and clustering (Fig.~\ref{CsigCFig}).
To better understand the effects of between-link and within-link correlations generated by article resampling,
we quantify and compare for the different resampling schemes the correlations between the weights of neighbor links and the variance of individual link weights.   
Our results show that between-link correlations of article resampling indeed are weak, but that the within-link correlations strongly affects the link-weight variance. 
Because multinomial resampling is almost as effective as article resampling, we propose a simple model that estimates the probabilities of multinomial resampling when full article-level data are not available.

\subsection*{Between-link correlations}
Article resampling introduces dependencies between link weights: an article may cite papers in different journals, so choosing that article adds citations to more than one journal simultaneously.
Here we want to measure how much these neighbor links are correlated in the resampled networks. 
 \begin{figure}[!ht]
\begin{center}
\includegraphics[width=0.5 \columnwidth]{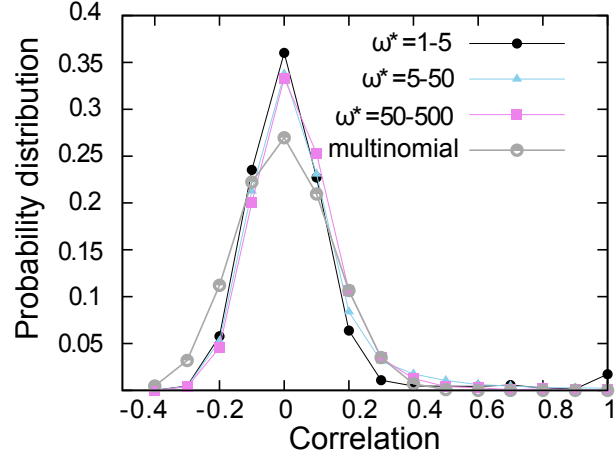}
\end{center}
\caption{
{\bf Neighbor links are only weakly dependent in article resampling.} The correlation distribution for a pair of neighbor links where at least one of them has a specific weight.
By definition, article resampling introduces correlation to the neighboring links and multinomial resampling ignores any correlation.
By comparing, we see that the result of correlation distribution confirms that most correlations of article resampling are not significant, when we compare them with the multinomial resampling as null mode. 
All points correspond to an average of at least 50 runs.}
\label{Corr}
\end{figure}
Figure \ref{Corr} shows that, in article resampling, only a fraction of neighbor links are weakly correlated. To check if these correlations are significant or not, we compare article resampling with multinomial resampling without between-link correlations. Figure \ref{Corr} confirms that between-link correlations are weak in article resampling.  
In fact, we could say that most neighbor links are not correlated, and that those few neighbor links that are correlated tend to be positively correlated. 
As we saw in the beginning of section  \emph{Results and Discussion}, this slight correlation doesn't have a great impact on the significant cluster cores or ranking of nodes. 
As shown, the dependency between links has a small effect on significant cluster cores, but nevertheless it influences the time that non-significant clusters emerge and can give a clue about important changes that might happen in the future. 

\subsection*{Within-link correlations}\label{SLVA}

\begin{figure}[!ht]
\begin{center}
\includegraphics[width=0.65 \columnwidth]{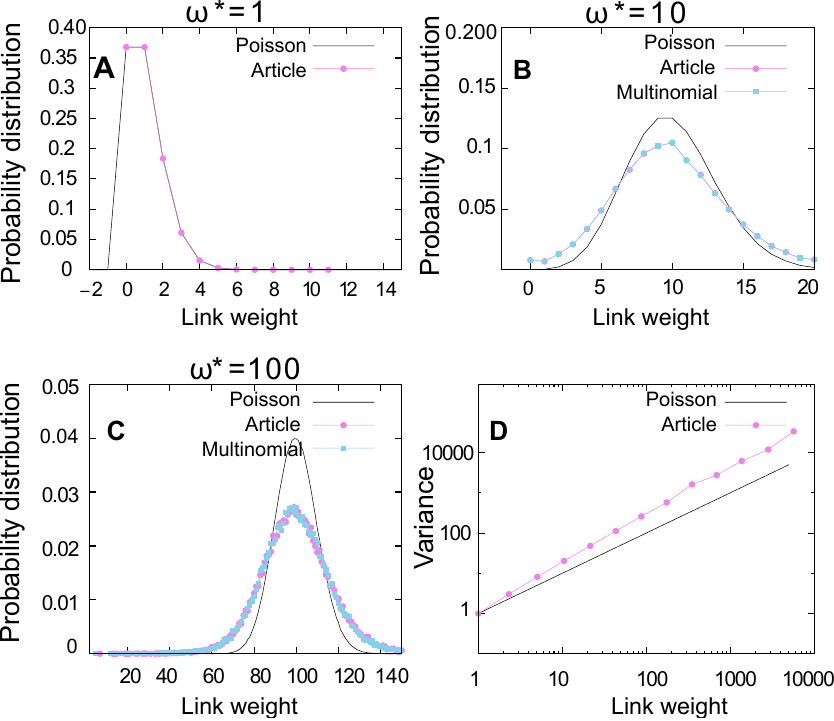}
\end{center}
\caption{
{\bf Comparing the probability distribution of link weights in article resampling with Poisson resampling and multinomial resampling.} \textbf{A} For low link weight ($w^*=1$), Poisson resampling precisely coincides with article resampling. \textbf{B,C} For medium values of link weight ($w^*=10$) and high values of link weight ($w^*=100$), Poisson resampling underestimates article resampling. The variance of the distribution in article resampling is much higher than in Poisson resampling. For example, for $w^*=10$, the variance in article resampling is $\sigma^2_{art}=19$, while the variance in Poisson resampling is $\sigma^2_{poiss}=10$. Similarly, for link weight $w^*=100$, the variance in article resampling is $\sigma^2_{art}=290$, while the variance in Poisson resampling is $\sigma^2_{poiss}=100$. The variance in multinomial resampling is quite close to article resampling, which confirms that the multinomial model imitates article resampling and make the distribution broader than Poisson resampling. \textbf{D} The variance of link weights in article resampling and Poisson resampling averaged over all resamples. All points correspond to averaging over 1000 runs.}
\label{Var}
\end{figure}

Figure \ref{Var} shows how much a specific link weight, $w^*$, varies based on the resampling scheme. 
When the link weight is very small, for example, $w^*=1$, we see that the variance of link weights in Poisson resampling perfectly matches with the variance in the article resampling (Fig.~\ref{Var}A).
The link weight equal to one means that only one article contributes to the citation between two journals, 
so the chance of picking that article is $\frac{1}{N}$, where $N$ is the total number of articles. Therefore, after resampling $N$ articles, the chance of getting that specific paper $k$ times is $\frac{1}{N}^k(\frac{N-1}{N})^{(N-k)}$, which, in the limit of large $N$, coincides with the definition of $Poisson(1)$. 
But when the link weight between two journals is higher than one, for example, medium values such as $w^*=10$ in Fig.~\ref{Var}B or high values such as $w^*=100$ in Fig.~\ref{Var}C, we see that the variance of Poisson resampling underestimates the variance of article resampling. This happens because citations can come in groups:
for a link where its weight $w^*$ is medium/high, there are $A$ articles ( $A<=w^*$) that contribute to that weight, and sometimes articles might add more than one citation.
So, although article resampling gives the same average weight as Poisson resampling, the variance of that weight in article resampling would be higher than for Poisson resampling. 
In summary, high link weights result in greater differences between the variance of article resampling and Poisson resampling (Fig.~\ref{Var}D).

Indeed, although Poisson resampling assumes an enormous number of binomial events that produce a specific link weight, article resampling tells us that the observed link weight is the outcome of multinomial events. 
In multinomial resampling, every link weight is generated from a multinomial distribution independently from other links. 
Although multinomial resampling assumes independency between links' weights and  article resampling does not, Fig.~\ref{Var}(B,C) shows that multinomial resampling completely matches  article resampling on the link level. \begin{figure}[!ht]
\begin{center}
\includegraphics[width=0.65 \columnwidth]{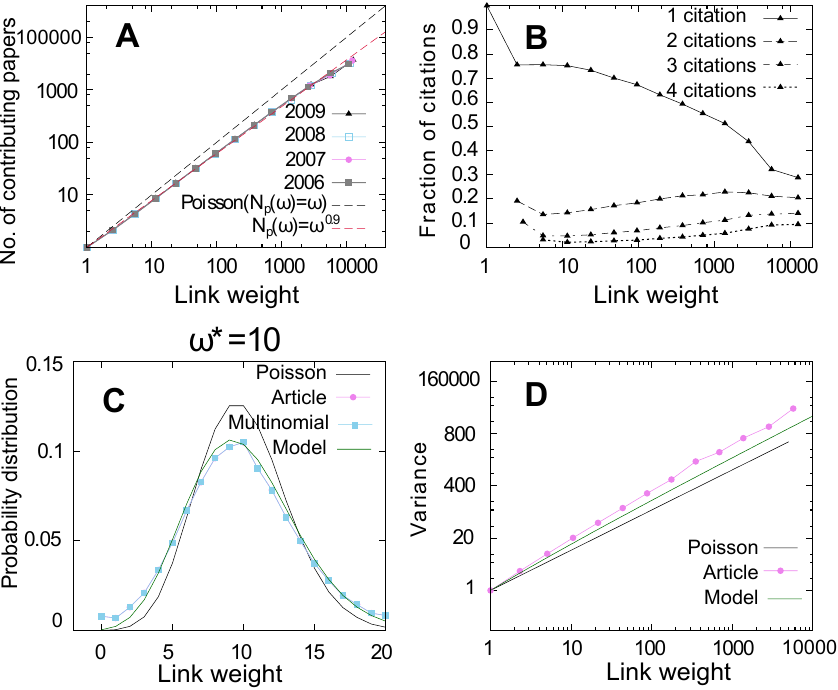}
\end{center}
\caption{
{\bf The high variance of article and multinomial resampling can be estimated by a simple model that extends Possion resampling to account for papers that contribute multiple citations to the same journal.} \textbf{A} Average number of papers that contribute to a specific link weight in logarithmic scale. For all years, the number of papers with weight $w$ ($N_P(w))$ fits to the function $x^\alpha$ with exponent $\alpha = 0.9$.  \textbf{B} The fraction of 1, 2, 3 and 4 citations that contribute to building a specific link weight $w$. Compared to low link weights, high link weights have a lower fraction of papers with only one citation and a higher fraction of papers with 2, 3 or 4 citations.  \textbf{C} The probability distribution of link weight $w^*=10$ for 4 cases: Poisson resampling, article resampling, multinomial resampling, and the minimal model. The high variance of article/multinomial resampling could be estimated by the model.  \textbf{D} The model can generate higher variance than Poisson resampling for different link weights. However, it could not generate exactly as high a variance as article resampling.}
\label{ContPaper}
\end{figure}

Multinomial resampling intrinsically considers group citations, and therefore it can generate higher variance than Poisson resampling. 

But what if the probabilities of different link weights are unknown for a given network? To estimate the probabilities, we look at the number of papers that contribute to a link with a specific weight. Figure \ref{ContPaper}A shows that, when the link weight $w$ is high, the number of papers that contribute to generating that link weight $N_P(w)$ is far from the value of the weight itself. 
Figure \ref{ContPaper}B shows that, when the link weight increases, the fraction of single citations that contribute to that weight is reduced. As Fig.~\ref{ContPaper}A shows, the number of papers that contribute to generating a link weight $w$ scales as $w^{0.9}$ for all years. We use this information to build a model for estimating the multinomial distribution when the probabilities of different link weights in a given network are not known. 
We assume that each weight $w$ is generated from papers with only one or two citations. We can simply estimate the number of papers with one citation $N_1$ and the number of contributing papers with two citations $N_2$ by solving the following linear equation system:  
\begin{equation}
\begin{split}
N_P(w)=N_1+N_2=w^{0.9}\\
N_1+2N_2=w
\end{split}
\bigg\}
\Rightarrow 
\begin{split}
N_1=2w^{0.9}-w\\
N_2=w-w^{0.9}
\end{split}
\end{equation}
After estimating $N_1$ and $N_2$, we suggest resampling every link weight by using the following minimal model:
\begin{equation}
Poisson(N_1)+2Poisson(N_2)
\end{equation}
The variance that we could get from this model is:
\begin{equation}
\begin{split}
Var(Poisson(N_1)+2Poisson(N_2))=\\
3w-2w^{0.9}
\end{split}
\end{equation}
In Fig.~\ref{ContPaper}C, we show the probability distribution of link weight $w^*=10$ for four cases: Poisson resampling, article resampling, multinomial resampling, and the proposed minimal model.  As shown, the high variance of article and multinomial resampling could be estimated by the minimal model. However, this estimation is not exact because the minimal model does not take into account group citations with three or more citations. In summary, the model can generate higher variance than Poisson resampling for different link weights, but it can not generate exactly as high a variance as article resampling (Fig.~\ref{ContPaper}D). 


\section* {Conclusion}
Link correlation of a resampling scheme influences the significance analysis of communities and ranking.
We compare three scenarios: fully maintained correlations between and within links (article resampling), no correlations between links (multinomial resampling), and completely broken link correlations (Poisson resampling).
We found that the result of significance analysis in multinomial resampling almost matches with article resampling. We conclude that the role of variance of individual links is greater than the role of correlation between links.
Nevertheless, we found that conserving link correlation in a resampling scheme can provide an early hint of possible changes to the network in the future.
The basic approach that we have laid out here, resampling the more or less independent components of a network for significance analysis, can be applied to other networks than citation networks. We speculate that the variance of link weights will play the major role also in those networks.
These findings can help researchers to better understand and asses reliable significant communities and structural changes for a given weighted network.





\section*{Acknowledgements}
We are grateful to Sara de Luna and Deborah Kolp for many valuable discussions. 
MR was supported by Swedish Research Council grant 2009-5344.



\end{document}